# Opportunities and challenges of mobile learning for promoting mathematical literacy

**Zaenal Abidin**
Faculty of Mathematics and Natural Sciences
Semarang State University
Semarang, Indonesia
Email: Z.Abidin@mail.unnes.ac.id

**Anuradha Mathrani**
School of Engineering and Advanced Technology
Massey University
Auckland, New Zealand
Email: A.S.Mathrani@massey.ac.nz

**David Parsons**
The Mind Lab, Unitec
Auckland, New Zealand
Email: david@themindlab.com

**Suriadi Suriadi**
School of Engineering and Advanced Technology
Massey University
Auckland, New Zealand
Email: S.Suriadi@massey.ac.nz

## Abstract

Mathematical literacy plays an important role in supporting individuals to fulfil their professional roles in modern society. The affordances of mobile technologies as well as the emergence of new theories in mobile learning have the potential to promote mathematical literacy. However, implementation of mobile learning in Indonesian society faces challenges related to perceived ethical and learning issues in curriculum-based educational settings. This study aims to investigate the preparedness of teachers in integrating mathematics subject content with mobile technologies, especially in promoting mathematical literacy. An exploratory study has been conducted using mixed methods. Questionnaire survey and semi-structured interviews were conducted to understand teachers' knowledge in mathematical literacy and to identify opportunities and challenges for mobile learning within instruction. Findings indicate that teachers mostly do not know about mathematical literacy, indicating that the concept of mathematical literacy needs to be promoted. Further, most schools prohibit the use of mobile devices in classrooms as they are wary of inappropriate use of mobile devices which may harm students' mental health and distract them from learning. Study finds this to be the most common cause for teachers' reluctance in using mobile learning.

## 1　INTRODUCTION

The term mathematical literacy has become quite well-known internationally through the Program for International Student Assessment (PISA) conducted by the Organisation for Economic Co-operation and Development (OECD) (De Lange 2006). Mathematical literacy is a subset of mathematical competencies, which means that to be mathematically competent, one must be mathematically literate and be able to resolve problems encountered in daily life (Niss 2015). The National Council of Teachers of Mathematics (NCTM) acknowledges that technology is an essential component of a learning environment; however, having access to technology needs to be complemented by teachers who act as mediators and who know how to use technological tools (NCTM 2011). Hence, to improve teaching and learning outcomes, a good understanding on how technology relates to the underpinning pedagogy and curriculum content is required (Koehler et al. 2007).

The proliferation of feature-rich mobile technologies as well as the emergence of new theories in mobile learning have raised a lot of attention on the way in which mobile technologies can transform and reconstruct educational practices (Crompton and Burke 2014). Study performed by Kalloo and Mohan (2011) reveals that there is a strong correlation between students' performance and the use of mobile learning applications. Kalloo and Mohan (2011) confirmed that students who used mobile applications more frequently and longer showed better performance than those who did not use them as frequently. However, the design of mathematical tasks and the implementation of mobile learning elements are equally important for improving students' competencies (Lee and Kautz 2015). Further, Lee and Kautz (2015) developed a framework for mobile learning task design and implementation to guide teachers in incorporating mobile technology into mathematics lesson. They highlight that the success of mobile learning experience is determined by prior experience, preparation, and skills of users both teachers and students.

However, in implementing mobile learning within instruction some ethical issues and concerns might arise. Many teachers have expressed concern about use of mobile devices in their classrooms as they may distract students from learning activities (Dyson et al. 2013; Keengwe et al. 2012). Consequently, many schools have banned the use of mobile devices within instruction.

The paper aims to capture the big picture of mobile phone usage in mathematics classrooms. An exploratory study has been conducted to investigate the preparedness of teachers in integrating mathematics subject content with mobile technologies. This study explores the following research questions:
RQ1. How many teachers currently understand about mathematical literacy?
RQ2. To what extent are teachers using mobile learning to teach mathematics in classrooms?
RQ3. What challenges do they face in implementing mobile learning?

## 2　BACKGROUND

### 2.1　ICT and Mathematics Education in Indonesia

The regulation (reference number 68 in the curriculum year 2013) of Ministry of Education and Culture (MoEC) has affirmed that all subject matters (including mathematics) are to be integrated with ICT and has mandated teachers to build the necessary skills to integrate technology into instruction (MoEC 2013). To this end, information and supports for mathematics teachers are required with regards to integrating technology in teaching mathematics (Lew and Jeong 2014). Many technological devices provide range of functionality, such as computational capabilities and rich graphical interfaces provided by many technological devices enable all sorts of rich functionality suitable for enhancing the process of teaching and learning mathematics (Niss et al. 2007). However, the ICT skill levels of the majority of teachers in Indonesia are still quite low (Copriady 2014). The result of UKG of Indonesia (the national examination of teachers' competency), conducted online in 2011 and 2012, also corroborates this evidence. The national average score for UKG in 2012 was 47.84/100, far from the passing grade of 70/100. This failure is not because teachers lack experiences in teaching; on the contrary, this is because teachers were not aware of the technicalities in dealing with online exams (Yusri and Goodwin 2013). Given this fact, ICT integration into curriculum in Indonesia is still far from the desired expectations.

Besides teachers' competency issues, Indonesia is now at a critical point of student performances especially in mathematics (Edo et al. 2014). An international comparative study like PISA 2012 shows that mathematics performance of Indonesian students are the second lowest in the league table and 75.7% of students were unable to reach level 2 (OECD 2014). The PISA survey indicates that Indonesian students have not been much mathematically literate. In addition, the average mathematics score of Indonesian students in the Trends in International Mathematics and Science Study (TIMSS) 2007 was ranked 36 out of 49 countries (The World Bank 2010). Responding to the results of these studies, the MoEC argued that the performances of Indonesian students were not satisfactory because the test materials in the PISA and TIMSS are not available in the Indonesian curriculum (MoEC 2013).

### 2.2　Mathematical Literacy

Identifying and developing mathematical literacy is an important aspect to improve the quality of instructional practice in the teaching of mathematics (Yavuz et al. 2013). Mathematical literacy supports an individual to become a professional in modern society (Stacey 2012). In particular, the definition of mathematical literacy for the purpose of PISA is

> An individual's capacity to formulate, employs, and interprets mathematics in a variety of contexts. It includes reasoning mathematically and using mathematical concepts, procedures, facts and tools to describe, explain and predict phenomena. It assists individuals to recognise the role that mathematics plays in the world and to make the well-founded judgments and decisions needed by constructive, engaged and reflective citizens (OECD 2013).

The definition clearly stresses the need to develop individuals' capacity in understanding and using mathematical content in real situations. The use of verbs "formulate", "employ" and "interpret" refer to the roles of individuals as active problem solvers (OECD 2013).

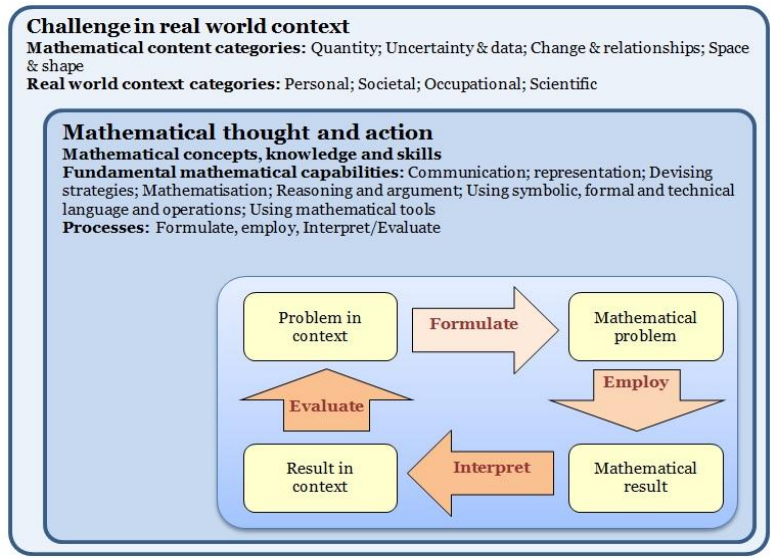

*Figure 1: A mathematical literacy model in practice (OECD 2013)*

The main points of the definition are suitable with the concept of "problem-solving process" or "mathematisation". Mathematisation cycle consists of five stages, namely starting with a contextualised problem in a real-world setting; employing mathematical concepts to organise the information obtained; representing the real-world problem mathematically into mathematical models; solving the mathematical problem; and evaluating the mathematical solution in the context of problem to determine whether the derived solution makes sense (De Lange 2006).

In practice, framework for mathematical literacy comprises various components as seen in Figure 1. The PISA's framework for mathematics is represented by ML + 3C. ML is an abbreviation for mathematical literacy, and the three Cs are abbreviation for content, context and competency (OECD 2003). The outer-most box in Figure 1 shows that mathematical literacy takes place in the context of challenge that arises in the real-world setting. In this framework, these challenges are characterised in two ways. The challenges or problems are characterised by the areas of life from which the problem arises and the nature of the mathematical phenomenon that underlies the challenge. The middle box shows mathematical thought and action such as mathematical concepts, knowledge and skills, fundamental mathematical capabilities and processes of actions in mathematical literacy. Moreover, the inner-most box delineates model of solving a real-world problem with mathematics (OECD 2013).

As an illustration, suppose a situation in a real-world setting involves estimating how much water it takes to fill up an entire swimming pool, the situation provides a context for the mathematical task. To solve the problem, a student needs knowledge of mathematical content (e.g., formula, unit of measurement). In order to solve the problem successfully, a student needs certain competencies (e.g., mathematising, using symbols and operations).

## 2.3  Mobile Learning

### 2.3.1  Definition of mobile learning

There have been various definitions to describe the term of mobile learning. For example, Park (2011) describes mobile learning as "the use of mobile or wireless devices for the purpose of learning while on the move". Further, O'Malley et al. (2005) has defined mobile learning as "any sort of learning that happens when the learner is not at a fixed, predetermined location, or learning that happens when the learner takes advantage of the learning opportunities offered by mobile technologies". Meanwhile, Sharples et al. (2007) define mobile learning as "the processes of coming to know through conversations across multiple contexts amongst people and personal interactive technologies". The variety of definitions shows dynamically evolving interest in this field as mobile technologies have progressed so the definitions too have changed over time, relying on the constructs used in these studies (Crompton 2013).

### 2.3.2 Opportunities and challenges of mobile learning

Mobile devices in this study comprise any handheld device capable of multiple functions, including but not limited to accessing the internet, running applications, listening to music, taking pictures and recording audio/video. Examples include mobile phone, smartphone, tablet or similar devices. Laptop or notebook is not included. A number of embedded features in mobile devices can be beneficial to teaching and learning activities (Thomas et al. 2013). Further, mobile devices can support communication, enhance collaboration between and among students and teachers, to promote a dynamic learner-centred learning environment (Aubusson et al. 2009; Looi et al. 2010). In addition, mobile devices can enhance learning within an authentic context and culture (Naismith et al. 2004). Mobile devices also enable learners to personalise their own learning tailored to their needs (Parsons 2014).

Although mobile learning has promising future potential, the opportunities for mobile learning do not come without challenges. Mobile phones in the classroom have been perceived as a disruptive technology for teachers (Thomas et al. 2013). A ringing phone is considered the most common disruption in the classroom and may negatively impact on student performance (Thomas et al. 2013). Teachers also have concerns about students' use of mobile phones for cheating and collusion in which students text answers during exams, take pictures of exam papers to share with friends, store answer keys to be consulted in exams, or find answer sources via the internet during exams (Dyson et al. 2013; Keengwe et al. 2012). Students also use their mobiles for inappropriate activities (sexting) (Thomas et al. 2013). Sexting can lead to harassment and cyberbullying (Siegle 2010). The consequences of cyberbullying can be dire. A number of teens have been reported to have committed suicide after being victims of cyberbullying (Keengwe et al. 2012). Given these facts, many schools have banned the use of mobile phones in classrooms since they can distract from learning. These challenges of mobile learning have been used in designing the questionnaire survey in this study to explore possibilities of mobile learning in mathematics instruction.

## 2.4 Convergence of Mobile Learning and Mathematical Literacy Practices

As mentioned above, mobile learning has potential for mathematics instruction. However, how mobile learning usage promotes mathematical literacy and what pedagogical approaches can facilitate learner to be mathematically literate has not been explored. Table 1 details how using mobile learning within instruction can assist students in acquiring mathematical literacy skills.

As implied in the definition of mathematical literacy, learners are supposed to be active problem solvers. Mobile learning supports dynamic learner-centred learning which enables a learner to take an active role in learning (Looi et al. 2010). Further, according to Vygotsky's theory (as cited in Leikin and Zaslavsky 1997) learners are able to solve certain problems by working together before they are ready to solve the same problems on their own. The value of learner collaboration is essential in effective learning of mathematics. Mobile learning further facilitates the process of collaborative learning (Aubusson et al. 2009). Mathematical literacy emphasize students' capacity in understanding and in using mathematical content in real situations; with respect to this, mobile learning facilitates situated learning which is promoting learning through activities within an authentic context (Naismith et al. 2004).

All information in this section shows that (1) there is a need to improve mathematical literacy of Indonesian students, and (2) mobile-assisted mathematical literacy have the potential to enhance students' capacity in understanding mathematics and apply them in real-world contexts – the essence of mathematical literacy.

However, before one implements the use of mobile devices in Indonesian schools, one needs to understand the feasibility and/or the need to do so. In particular, in order for this endeavour to be successful, at least two factors must be in place: firstly, the teachers themselves must be aware of what is mathematical literacy is all about, and secondly, the use of mobile devices within classrooms must be something that is actionable within the community. The research that we have conducted attempts to answer those questions. The details and the results of our study are explained in the remainder of this paper.

| PISA's capabilities | Mathematical literacy | Mobile learning |
|---|---|---|
| Communication | Capability to read, decode and interpret statements, ask questions, plan tasks, or use objects to form mental model of situations. Be able to give an explanation or justification by presenting the solution either during or after the process of finding a mathematical solution. | Mobile device features like camera, audio/video recorders and wireless connection enable learners to do synchronous/asynchronous communication, make a video or take a picture to explain and justify the solution of mathematics. |
| Mathematising | Transforming a contextualised problem to mathematical form, or interpreting or evaluating a mathematical outcome in the context of the original problem. | The outcome of mathematising process is a mathematics model. The model can be solved with mobile applications. |
| Representation | Presenting mathematical objects and situations such as formulae, tables, diagrams, charts, matrices, and concrete materials to describe situations or simplify forms of mathematical solution in order to make interpretation easily. | Mobile devices provide functions for representation of mathematical objects. A variety of mobile applications can be used to represent graphs, tables, diagrams, pictures, and equations. |
| Reasoning and argument | Thinking logically to explore and link elements of a problem to make a good decision as a part of the inference process. Also, making and checking justification of solutions to problems. | The learner can think logically from different types of mathematics problems by going over tutorials and quizzes provided on the mobile device. |
| Devising strategies for solving problems | Solving problems mathematically by selecting and devising strategy to get solutions in the context of the given problem. | After sufficient practice through tutorials and quizzes provided on the mobile device, a learner is able to make judgment on the best way to solve mathematics problems. |
| Using symbolic, formal and technical language and operations | This capability is related to understanding, interpreting, manipulating, and making use of mathematics symbols governed by mathematical conventions and rules. | Understanding of symbols and utilising formal constructs can be enhanced by text-based instructions provided in the mobile device, or by viewing video-based examples of the problem being solved. |
| Using mathematical tools | Mathematical tools provide apparatuses or media to facilitate individuals in the process of deriving solutions mathematically. Individual is required to be familiar with various kinds of mathematical tools. | Mobile devices with some mathematics apps provide the apparatuses to support instruction in classrooms. To take advantage of mobile devices, the learner must be familiar with features and the apps being used. |

*Table 1. Relation of fundamental mathematical capabilities with mobile learning*

## 3　METHODOLOGY

### 3.1　Procedure

The purpose of this study is to examine teacher's knowledge towards mathematical literacy and to identify opportunities and challenges of mobile learning within instruction to promote mathematical literacy. The study employed a mixed methods approach, i.e. a procedure for collecting, analysing, and integrating both quantitative and qualitative methods to understand a problem (Creswell et al. 2004). In this study, quantitative method was conducted through a survey that can be subjected to statistical analysis. The questionnaire used in the survey consisted of 20 items including both closed and open-ended questions. Further, qualitative data was performed through open-ended questions on the survey and through semi-structured interviews. In the interviews, teachers were asked to give their experiences in using mobile devices in teaching and learning activities and the challenges they face when mobile learning is implemented in their school. The interviews were conducted in Indonesian language and were recorded with the consent of interviewees.

### 3.2　Respondents

This study was conducted in Semarang municipality, Central Java province, Indonesia. First, a survey was carried out in April 2015. In this survey, the sample size is determined by a formula proposed by Yamane (1967). The number of mathematics teachers in Semarang municipality is 462 teachers. Therefore, by using Yamane's formula with a margin of error of 0.05 the sample size is 214 teachers. In this study, 213 teachers from 129 different junior high schools participated, which is, very close to the target number. Teacher participants were predominantly female, with 61.5% (131) and 38.5% (82) male. Further, most of teachers were from the urban area (195 or 91.5%), with very few (18 or 8.5%) from the rural area. Moreover, most of teachers (157 or 73.7%) had teacher's certificate from the teacher certification program while the rest (56 or 26.3%) did not have.

Next, interviews were held with ten teachers from different schools in May 2015. These teachers had participated in the initial survey. Each interview took approximately 20 minutes. Four teachers were interviewed face to face at the concerned teacher's school while six teachers could not be met face to face since they were busy in preparation of national exam for their schools, so they were interviewed over phone. In this paper, the teachers are identified by their user codes (R1, R2, etc.)

### 3.3　Analytical Strategy

Quantitative data gained from the survey was first coded and then analysed using descriptive statistics functionality available in SPSS (IBM 2015). The outcomes have been used to understand the experiences of respondents in their use of mobile devices in teaching and learning activities, and also to gauge respondents' knowledge about mathematical literacy. Meanwhile, qualitative data obtained from the interview were transcribed and analysed in Indonesian language. The transcribed data has been analysed to identify categories by dividing each type of the gathered data into segments and examining these segments for similarities and differences. Next, each response was coded to a number of categories. After coding the responses, the categories which had the most responses were marked as prominent. The next step is to see which categories are related and where patterns and trends can be identified. Once the more prominent categories were identified and selected, they are used to connect the rest of the categories and to build conceptual model of the studied phenomenon. At this step, relevant quotations were translated into English, and were classified based on the categories along with quotation numbers from each interviewee, so that it would be easier to trace issues to the original transcripts later on.

## 4　RESULTS

This section presents the results based on the analysis of the questionnaire results and interviews. We present the results in three subsections; teachers' knowledge about mathematical literacy (RQ1), teachers' experience to mobile learning usage in mathematics class (RQ2), and teachers' perceived challenges of using mobile learning (RQ3).

### 4.1　Teachers' Knowledge about Mathematical Literacy

This subsection presents how many teachers have knowledge in regard to mathematical literacy (RQ1). Most (137 or 64.3%) of teachers do not know the term mathematical literacy. From those teachers who knew about mathematical literacy, 20.7% (44) of teachers got mathematical literacy information from the internet, 12.2% (26) from workshop/teacher's training, 6.6% (14) from journal articles, and 2.8% (6) from other sources such as lectures, theses, mathematical literacy contests, and books. Regarding the mathematical literacy contest held each year, only few teachers (9 or 4.2%) from eight different schools had sent their students to participate. These facts indicate the lack of appreciation towards such contests from the teachers and the schools. One teacher said that her school sends students to mathematics Olympiad since this competition is considered more popular than mathematical literacy contest.

*"We have never participated in mathematical literacy contest, but each year we send our students to regional mathematics Olympiad. Being a teacher myself, I have never given students tasks related to*

*mathematical literacy. I am not familiar with such tasks, sometimes, I give them tasks related to real-world context during learning activities*" (R9).

Another teacher said that he is less familiar with mathematical literacy tasks because he does not understand the concept.

"*I still do not understand how* [mathematics] *tasks are characterised by mathematical literacy. So, I am less familiar with it*" (R2).

These facts show that dissemination of mathematical literacy to mathematics teachers needs to be widened.

### 4.2　Teachers' Experience to Mobile Learning Usage in Mathematics Class

Before we asked the teachers about their experiences in using mobile devices for teaching and learning activities (RQ2), we enquired how many teachers currently use mobile devices. Over a half of teachers owned smartphones (124 or 58.22%), tablet (28 or 13.15%), and internet-enabled basic phone (23 or 10.80%). Meanwhile, some (40 or 18.78%) of teachers do not have mobile phone with internet capabilities. Further, we enquired about the use of mobile devices in their daily activities. Almost half of teachers reported that they mostly use mobile devices for texting or sending messages (104 or 48.83%), getting information (62 or 29.11%), accessing social network (51 or 23.94%), and checking emails (35 or 16.43%). Other activities such as listening to music or watching video, playing games, getting direction, and reading content (e.g., e-book, article, etc.) were less frequent (see Figure 2 for specific results).

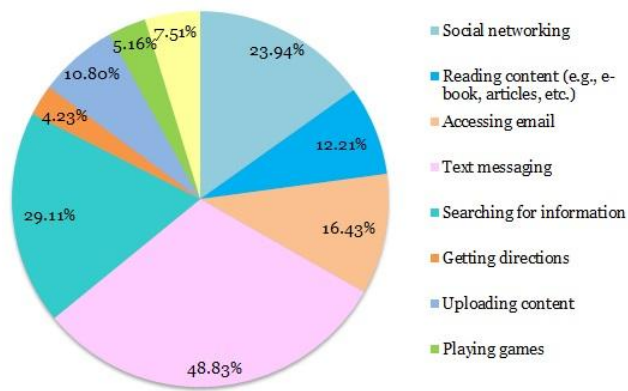

*Figure 2: Teachers' activities in the use of mobile devices*

Next, we asked teachers about their experience in using mobile devices for mathematics instruction. The survey reveals that only 31.9% (68) of teachers have used mobile devices in teaching and learning activities. They use mobile learning in sorts of activities either indoor (28 or 13.1%) or outdoor (19 or 8.9%) activities as well as either in formal (24 or 11.3%) or informal (3 or 1.4%) setting.

With respect to the use of mobile phone in mathematics class, a teacher shared their experiences.

"*I use the Quipper School application* [a web-based mobile application]. *I put assignments in the Quipper School and ask students to complete them within one-week time at home. During this time, the results can be tracked. Students with low, high, and medium level of mathematical ability can be recognized immediately*" (R10).

This view indicates that respondent R10 perceives mobile phone usage in instruction is beneficial. The affordances of mobile technology enable teachers to enrich learning experience, creating innovation learning, and expand horizons in adoption of technology for teaching and learning activities.

### 4.3　Teachers' Perceived Challenges of Using Mobile Learning

Teachers were asked to identify the challenges in use of mobile learning in mathematics instruction (RQ3). Based on the survey, 58.2% (124) of teachers responded that mobiles will disrupt the class (e.g. phones ringing during class, texting and checking incoming phone message in classroom). Second challenge voiced is that of cyberbullying and sexting (106 or 49.8%), followed by cheating (87 or 40.8%). These issues have made schools establish policies to ban mobile devices in classrooms. To deal with these ethical issues, two teachers shared their recommendations.

"*…we need to invite representatives from school committees, parents, school counsellors, and teachers to discuss about this issue so that we can adjust the regulations governing the use of mobile learning in teaching activities*" (R4).

"*My solution is that if the students want to use mobile technology for learning which, I believe is good, then we first have to consult and coordinate with*-[stakeholders, including] *the homeroom teachers, the student council adviser, and most importantly, the principal… After the stakeholders agree with what have been discussed, we need to inform the students' parents … we* [should] *send official letters to them stating that on this day and at this time, and so on, students are allowed to, or maybe must, bring their mobile phones for learning purposes*" (R6).

From these recommendations, we can see that both teachers share similar views that mobile learning can still be implemented in teaching and learning activities by involving all stakeholders, and by modifying schools' policy regarding mobile devices restriction. Furthermore, the readiness of teachers towards mobile learning is also a decisive in the success of this endeavor. One teacher acknowledged this view.

"*Firstly, teachers must be prepared. Be prepared,* [I] *mean teachers must further improve their teaching materials preparation skills* [using mobile learning facility]. *Secondly, teachers should be able to monitor their students, and the students' learning process must be interesting*" (R10).

Apart from teachers' concern about using mobile devices within instruction, findings show that limited availability of technology (e.g. students do not have mobile devices, schools have insufficient / do not have mobile devices) is also a barrier in the implementation of mobile learning. When we asked about this issue, some (70 or 34.3%) of teachers answer yes. From the interview, one teacher confirmed this issue.

"*…the challenge* [of mobile learning] *is that not all students have an android phone or a smartphone*" (R8).

Another challenge perceived by teachers is the wireless connectivity availability. Wi-Fi enables new style of instructions to emerge because access to the Internet becomes ubiquitous. However, not all schools have good Wi-Fi connection and even so, the connection only covers a certain area. Two teachers voiced this issue:

"[Wi-Fi signal] *does not cover the whole* [school] *area, but is only available within a certain area. When we need it, we sometimes need to move the class*" (R3).

"*…the second challenge is the overload of Wi-Fi networks usage leading to slower access time in loading resources* [from the Internet]" (R5).

In this regards, it seems that wireless connectivity is crucial in the implementation of mobile learning. No Wi-Fi connectivity means that there is no Internet access on mobile devices. Although, there are various Internet data packs offer by companies, the mobile learning implementation would be much costly.

## 5   DISCUSSION AND RECOMENDATIONS

This study provides interesting insights on how mathematics teachers view integration of mobile technology in instruction for promoting mathematical literacy. Findings indicate that majority of teachers do not know much about mathematical literacy concepts. Also teacher participation in the mathematical literacy workshop is very low and, mathematical literacy contest is less popular than other student competitions like mathematics Olympiad. This is understandable since mathematical literacy concept is different from Indonesia's mathematics curriculum. These could be the factors contributing to the prevailing lukewarm interests from mathematics teachers in activities related to mathematical literacy. However, further investigation is needed to find out the cause.

Most teachers reported that they have owned mobile phones but many of them never use their mobiles for teaching and learning activities. Teachers use mobile phones for some activities, such as social networking, reading contents (e.g., e-book, articles, etc.), accessing email, texting, searching information, getting direction, uploading content, and playing games. These indicate that teachers do have technological skills. However, simply having technological skills is not enough to integrate technology into instruction. Teachers need to know the pedagogical role of technology within instruction (NCTM 2011).

Adopting mobile technology in education raises some ethical issues and concerns (Dyson et al. 2013). This study examined ethical considerations that might arise when students bring mobile phone in classrooms, such as distraction in learning, cyberbullying, sexting and cheating. More than a half of teachers perceived that mobile phone can be a distraction in classrooms and almost half were concerned about cyberbullying and sexting. Some of teachers were concerned about cheating. Lack of technology resources also were seen by teachers as inhibiting the use of mobile technology. More than one-fourth of teachers confirmed this matter. These issues are indeed the most-cited reasons causing teachers to be reluctant to use mobile learning within instruction.

These challenges need to be addressed if mobile technology will be used more effectively by teachers and their students. Regarding technology availability problems, Thomas et al. (2013) recommends that teachers should allow students who have mobile phones to work collaboratively with those who do not, and that school could procure mobile phones to facilitate students use in classrooms. Whereas, to deal with ethical concerns, a responsible mobile-use policy (RMUP) formula from Dyson et al. (2013) can be used to foster a sense of ethical personal responsibility on the part of mobile user. The RMUP sets four principles, namely (1) *enhanced learner agency*, the policy recognizes the key role of mobile technology in supporting learners in all aspects of their lives as well as acknowledge the value of mobile learning in supporting them to produce information for their learning purposes, (2) *responsibility* indicates the role of policy to encourage learners to be well behaved in using mobile technology, (3) *involvement of all stakeholders* refers all those who will be affected by the policy as well as those who will enforce it should be taking part in creating the policy, and (4) *focus on ethical behaviour* emphasises focus on the matter of real concern, which is how mobile technology will be used in learning.

# 6 CONCLUSION AND FUTURE RESEARCH DIRECTION

The results of this study are not without some limitations. The participants in this study were mathematics teachers who attended at a one-day workshop conducted by the department of computer science, Semarang State University in cooperation with the MGMP (secondary subject teacher forum) of mathematics of Semarang district, thus, were a convenience sample. Because this study was limited to teachers who attended the workshop, it does not give attention to teachers who did not attend, which limits the generalizability of the study.

The findings of this study provide perception of mathematics teachers about the use of mobile devices in the classroom to determine whether mobile technology can be used to promote mathematical literacy. Findings show that mobile learning has potential to improve students' performance to be mathematically literate. However, mobile devices, such as smartphones, mobile phones, and tablets are unlikely to be effective for use within instruction since many schools prohibit the use of mobiles in classrooms. Most teachers also perceived mobile devices as a disruptive technology. Moreover, many teachers confirmed that schools have insufficient facility with regard to mobile technology. Further, limited access of mobile devices in schools constitutes the main hindrance factor for process of adoption of mobile technology. These make teachers reluctant to use mobile learning in teaching and learning activities.

Finally, the findings indicate using alternative technologies that fit better with the current situation in Indonesia. This implies identifying teachers' skills with technology usage. Appropriate teacher-training programs can help teachers in improving their technology skills and in manifesting pedagogical knowledge with technology. Within this context, suitable training relevant to educational goal and priorities, knowledge and skills must be imparted. Teacher-training can be part of teacher professionalism development to improve quality teaching with technology. Teacher development should be ongoing to align with current standards and assessment methods. The next stage of the study will involve building on existing findings to define a framework which encompasses mobile technology as a mediator tool for mathematical literacy, and teacher learning. The framework can be applied for development of teachers' skills for implementing mobile literacy in mathematics classrooms.

## Copyright